\documentclass{PoS}

\title{Long-range Correlations in Massive Jets}

\ShortTitle{Correlations in Massive Jets}

\author{\speaker{Karoly Urmossy}\thanks{These research results were sponsored by the Chinese Academy of Sciences.}\\
        Research Institute for High-energy Physics of the Chinese Academy of Sciences\\
        19b Yuquan Road, Shijingshan District, Beijing China\\
        E-mail: \email{karoly.uermoessy@cern.ch; karoly@ihep.ac.cn}}

\abstract{We calculate the azimuthal anisotropy $v_2$ extracted from the large $\Delta y$ region of two particle $\Delta y-\Delta\phi$ correlations in a two-jet system, in which, the masses of the jets are not negligible compared to their energies. As the virtualities of the leading partons, initiating these jets are not negligible either, we use a recently developed, off-shell fragmentation model for the description of hadron production in the jets. We present the effect of the variation of jet mass and hadron multiplicity on the shape of the $v_2$ curve, and reproduce the low-multiplicity data set measured in proton-proton collisions at $\sqrt s = 13$ TeV.}

\FullConference{XXVII International Workshop on Deep-Inelastic Scattering and Related Subjects - DIS2019\\
		8-12 April, 2019\\
		Torino, Italy}

\usepackage{epsfig}

\usepackage{amssymb}
\usepackage{amsmath}
\usepackage{amsfonts}
\usepackage{amsbsy}
\usepackage{amscd}
\usepackage{amsopn}
\usepackage{amstext}
\usepackage{amsxtra}
\usepackage{graphicx}
\usepackage{color}
\usepackage{slashed}
		
\newcommand{\be}{\begin{equation}}
\newcommand{\ee}[1]{\label{#1} \end{equation}}
\newcommand{\ba}{\begin{eqnarray}}
\newcommand{\ea}[1]{\label{#1} \end{eqnarray}}
\newcommand{\nl}{\nonumber \\}

\begin{document}

\section{Introduction}
\label{sec:intro}
As has recently been observed \cite{bib:CMSv2pp,bib:CMSv2AA}, some long-range hadronic correlations are present in proton-proton (pp), proton-lead (pPb) and lead-lead (PbPb) collisions, resulting in final states with high hadron multiplicity, while these correlations are missing from low-multiplicity collisions of the same nuclei. More precisely, examinig the correlation function of hadron pairs as a function of their azimuth angle ($\Delta\varphi$) and rapidity ($\Delta y$) differences, one can realize that a ``ridge-like'' structure stretches along the $\Delta y$ axis around $\Delta\varphi=0$ (near-side region) when the number of hadrons created in the collision is high (around 200 or more), whereas, this ``near-side ridge'' is not present when the number of hadrons is no more than 20. The appearance of the near-side ridge might be the sign of the creation and expansion of a hot, dense quark-gluon plasma (QGP), and if we were able to cleanse the correlation function from all the effects \textit{not related to QGP}, we could make predictions on the initial geometry (impact parameter, nuclear overlap, etc...) of the collision, which induces pressure gradients that drive the expansion of the QGP. In this paper, we focus on a certain type of contaminating effect: correlations in a typically non-QGP related situation, a two-jet finalstate, in which, the initial partons are highly virtual, thus, there is enough phasespace in the initiated jets for final state particles with large azymuthal angle and rapidity separation. For the description of the hadronisation of the highly-virtual initial partons, we use the off-shell fragmentation model \cite{bib:UK} sketched in Sec.~\ref{sec:off}. We present the calculation of the $\Delta y - \Delta\varphi$ correlations and $v_2$ for two-jet events in Sec.~\ref{sec:cor}, and compare our results with experimental data in Sec.~\ref{sec:res}.

\section{Off-shell fragmentation}
\label{sec:off}
Following \cite{bib:UK}, we use the microcanonical statistical ensemble for the calculation of the distribution of hadrons stemming from a jet of momentum $P_\mu$ at an initial fragmentation scale, which we choose to be $Q_0 = \sqrt{P^2}$. As the phasespace of $n$ (massless) hadrons of total fourmomentum $P_\mu$ is 
\be
\Omega_n\left(P^\mu\right) \;=\; \prod_{i=1}^n \int \frac{d^3\mathbf{p}_i}{p_i^0} \delta\left(\sum_{j=1}^n p_j^\mu - P^\mu \right) \;\propto\; \left(P^2\right)^{n-2}\;,
\ee{off1}
the distribution of $n'$ hadrons stemming from the jet reads
\be
d_{n,n'}\left(p_1,\dots,p_{n'}, P\right)\,=\, \frac{\Omega_{n-n'}(P^\mu-p_1^\mu\dots-p_{n'}^\mu)}{\Omega_n(P^\mu)} \,\sim\, \left(1 - \sum_{i=1}^{n'}x_i + \sum_{i<j}^{n'}X_{ij} \right)^{n-n'-2}
\ee{off2}
with $x_{i} = 2P_\mu p_i^\mu/P^2$ and $X_{ij} = 2p_i^\mu p_{j\mu}/P^2$. As an example, the one-, and two-particle distributions are
\be
d_{n,1}\left(p_1, P\right)\,=\,  \frac{(n-1)(n-2)}{\pi P^2}\left(1 - x_1 \right)^{n-3}  \;,
\ee{off3}
and
\be
d_{n,2}\left(p_1,p_2, P\right)\,=\, \frac{(n-1)(n-2)^2(n-3)}{\pi^2 P^4}\left(1 - x_1 - x_2 + X_{12}  \right)^{n-4} \;,
\ee{off4}
with prefactors obtained from the normalisation condition $\prod_{i=1}^{n'} \int \frac{d^3\mathbf{p}_i}{p_i^0} \, d_{n,n'}\left(p_1,\dots,p_{n'}, P\right) = 1$.
Using the experimental observation that the hadron multiplicity in jets fluctuates according to the negative-binomial distribution (NBD)   
\be
\mathcal{P}_n \,=\, {{n+r-1}\choose{r-1}}\bar{p}^n (1-\bar{p})^r  \;,
\ee{off5}
the multiplicity averaged hadron distribution (fragmentation function FF) becomes 
\be
d_1\left(p_1, P\right) \,=\,  \sum_n \mathcal{P}_n\, n\, d_{n,1}\left(p_1, P\right)
\,=\,  A_1 \left(1 + \frac{q-1}{\tau}\,x_1\right)^{-1/(q-1)} - A_2\;.
\ee{off6} 
The parameters of the multiplicity distribution Eq.~(\ref{off5}) and those of the FF in Eq.~(\ref{off6}) are related as $r=1/(q-1)-3$ and $\bar p = (q-1)/(\tau +q-1)$. Eq.~(\ref{off6}) (often called the Tsallis-Pareto--distribution \cite{bib:parvan,bib:parvan2,bib:ishi}) describes hadron spectra measured in various high-energy collisions \cite{bib:wang}-\cite{bib:wang2}.

Using an appriximation, in which, the shape of the FF remains unchanged throughout the scale evolution, the dependence of its parameters on the fragmentation scale $Q=\sqrt{P^2}$ can be obtained: 
\ba
r(t) &\,=\,& \left[\frac{r_0+1}{r_0}\left(\frac{t}{t_0}\right)^{a} -1 \right]^{-1}\;,\nl
\bar{p}(t) &\,=\,& \left[1 + \frac{\frac{1- \bar{p}_0}{r_0 \bar{p}_0}\left(\frac{t}{t_0}\right)^{b} }{\frac{r_0+1}{r_0}\left(\frac{t}{t_0}\right)^{a} -1 }   \right]^{-1}\;,
\ea{off7}
where $t=\ln(P^2/\Lambda^2)$ and $t_0=\ln(P_0^2/\Lambda^2)$ with starting scale $P_0^2$, and $\Lambda$ being the scale where the 1-loop coupling of the $\phi^3$ theory $g^2 = 1/t\beta_0$ blows up. The additional parameters in Eq.~(\ref{off7}) are $a = [\tilde{\Pi}(3) + \tilde{\Pi}(1)]/\beta_0$ and $b = \tilde{\Pi}(3)/\beta_0$ with $\tilde{\Pi}(\omega) = 1/(\omega+1)(\omega+2)-1/12$, being the Mellin-transform of 
the 1-loop splitting function in the $\phi^3$ theory.

\section{Correlations in 2-jet events}
\label{sec:cor}
In this section, we examine 2-hadron correlations in events with two jets of momenta $P_1$ and $P_2$ in the final state with $N$ hadrons in total. Assuming that the multiplicity distribution and fragmentation function in one jet are independent from those in the other jet, the 2-particle distribution in a 2-jet event becomes, 
\ba
p_1^0 p_2^0 \frac{dN^{same}}{d^3\mathbf{p}_1 d^3\mathbf{p}_2} &\,=\,& \sum_{n=n_0}^N \mathcal{P}_n(P_1) \mathcal{P}_{N-n}(P_2) \left[ d_{n,2}\left(p_1,p_2, P_1\right) + d_{n,1}\left(p_1, P_1\right)d_{N-n,1}\left(p_2, P_2\right) \right. \nl
&& \qquad \left. d_{n,1}\left(p_2, P_1\right)d_{N-n,1}\left(p_1, P_2\right) +  d_{N-n,2}\left(p_1,p_2, P_2\right) \right]\;,
\ea{cor1}
with $n_0$ being some minimal number of hadrons that need to be produced within a jet. The superscript ``same'' denotes that both hadrons stem from the same event, whereas, the distribution of hadron pairs that stem from two different events is called ``mixed'': 
\ba
p_1^0 p_2^0 \frac{dN^{mixed}}{d^3\mathbf{p}_1 d^3\mathbf{p}_2} &\,=\,& \sum_{n_1=n_0}^N \mathcal{P}_{n_1}(P_1) \mathcal{P}_{N-n_1}(P_2)  \left[  d_{n_1,1}\left(p_1, P_1\right) +  d_{N-n_1,1}\left(p_2, P_2\right) \right] \times \nl
&& \times \; \sum_{n_2=n_0}^N \mathcal{P}_{n_2}(P_1) \mathcal{P}_{N-n_2}(P_2) \left[ d_{n_2,1}\left(p_2, P_1\right) + d_{N-n_2,1}\left(p_1, P_2\right)  \right]\;.
\ea{cor2}

We parametrize the momenta of the hadrons ($p_1$ and $p_2$) and jets ($P_1$ and $P_2$) as
\ba
p_{1,2} &\,=\,& p_T^{1,2} \left[ \cosh\left(y_{1,2} \right), \sinh\left(y_{1,2} \right), \cos\left(\varphi_{1,2} \right), \sin\left(\varphi_{1,2} \right)\right]\;,\nl
P_{1,2} &\,=\,&  \left[M_T^{1,2} \cosh\left(Y_{1,2} \right), M_T^{1,2} \sinh\left(Y_{1,2} \right), P_T^{1,2}\cos\left(\Phi_{1,2}\right), P_T^{1,2}\sin\left(\Phi_{1,2}\right)\right]\;,
\ea{cor3}
with jet transverse masses $M^{1,2}_T=\sqrt{M^2_{1,2} + P^2_{T1,2}}$ and jet masses $M_{1,2}=\sqrt{P^2_{1,2}}$. Writing the rapidities and azimuth angles of hadrons as $y_{1,2} = (Y\pm\Delta y)/2$ and $\varphi_{1,2} = (\Phi\pm\Delta\varphi)/2$, the energy fractions $x_{1,2}$ and $X_{12}$, being the arguments of the distributions (\ref{off3})-(\ref{off4}) appearing in Eqs.~(\ref{cor1})-(\ref{cor2}) become
\ba
x_{1,2}^a &\,=\,& \frac{2p^{1,2}_\mu P_a^\mu}{M^2_a} =  \frac{2p_T^{1,2}}{M^2_a} \left[M_{Ta} \cosh\left(\frac{Y\pm\Delta y}{2} - Y_a\right) - P_{Ta}\cos\left(\frac{\Phi\pm\Delta\varphi}{2} - \Phi_a\right)\right]\;,\nl
X_{12}^a &\,=\,& \frac{2p^1_\mu p_2^\mu}{M^2_a} =  \frac{2p^1_T p^2_T}{M^2_a} \left[\cosh\left(\Delta y\right) - \cos\left(\Delta\varphi\right)\right]\;,\qquad a=\lbrace1,2\rbrace\;.
\ea{cor4}
Using the above parametrisation, the signal (S), the background (B) and the correlations (C) are constracted from the ``same event'' and the ``mixed event'' distributions as  
\ba
S\left(p_T^1,p_T^2,\Delta y,\Delta\varphi\right) &\,=\,& \int dY \int d\Phi\; p_1^0 p_2^0 \frac{dN^{same}}{d^3\mathbf{p}_1 d^3\mathbf{p}_2}\;,\nl
B\left(p_T^1,p_T^2,\Delta y,\Delta\varphi\right) &\,=\,& \int dY \int d\Phi\; p_1^0 p_2^0 \frac{dN^{mixed}}{d^3\mathbf{p}_1 d^3\mathbf{p}_2}  \;,\nl
C\left(p_T^1,p_T^2,\Delta y,\Delta\varphi\right) &\,=\,& \frac{S\left(p_T^1,p_T^2,\Delta y,\Delta\varphi\right)}{B\left(p_T^1,p_T^2,\Delta y,\Delta\varphi\right)}\;.
\ea{cor5}
Finally, the azymuthal anisotropy is defined by the Fourier components of C:
\be
v_{n}\left(p_T,\Delta y\right) \,=\, \sqrt{ \frac{\int d\varphi \cos(n\Delta\varphi)\, C\left(p_T,p_T,\Delta y,\Delta\varphi\right)}{\int d\varphi\, C\left(p_T,p_T,\Delta y,\Delta\varphi\right)} }\;.
\ee{cor6}

\section{$v_2$ from long-range correlations}
\label{sec:res}
In this Sec., we calculate $v_2(p_T,\Delta y)$ given in Eq.~(\ref{cor6}) averaged over the rapidity range of $|\Delta y| \in [2,4]$ in a final state with two back-to-back jets in the transverse direction ($Y_{1,2}=0$) of momenta $P_{1,2} = \left(\sqrt{M^2_{1,2}+P^2_{jet}},0,\pm P_{JET},0\right)$. The magnitudes of the threemomenta of the jets are taken to be $P_{JET} = 40$ GeV/c in each cases, whereas, the jet masses and numbers of hadrons in the jets are varied. We compare the calculated results with data, measured in pp collisions at $\sqrt s =$ 13 TeV with low (10-20) and high (105-150) hadron multiplicity \cite{bib:CMSv2pp}.

\begin{figure}
\begin{center}
\includegraphics[width=0.48\textwidth]{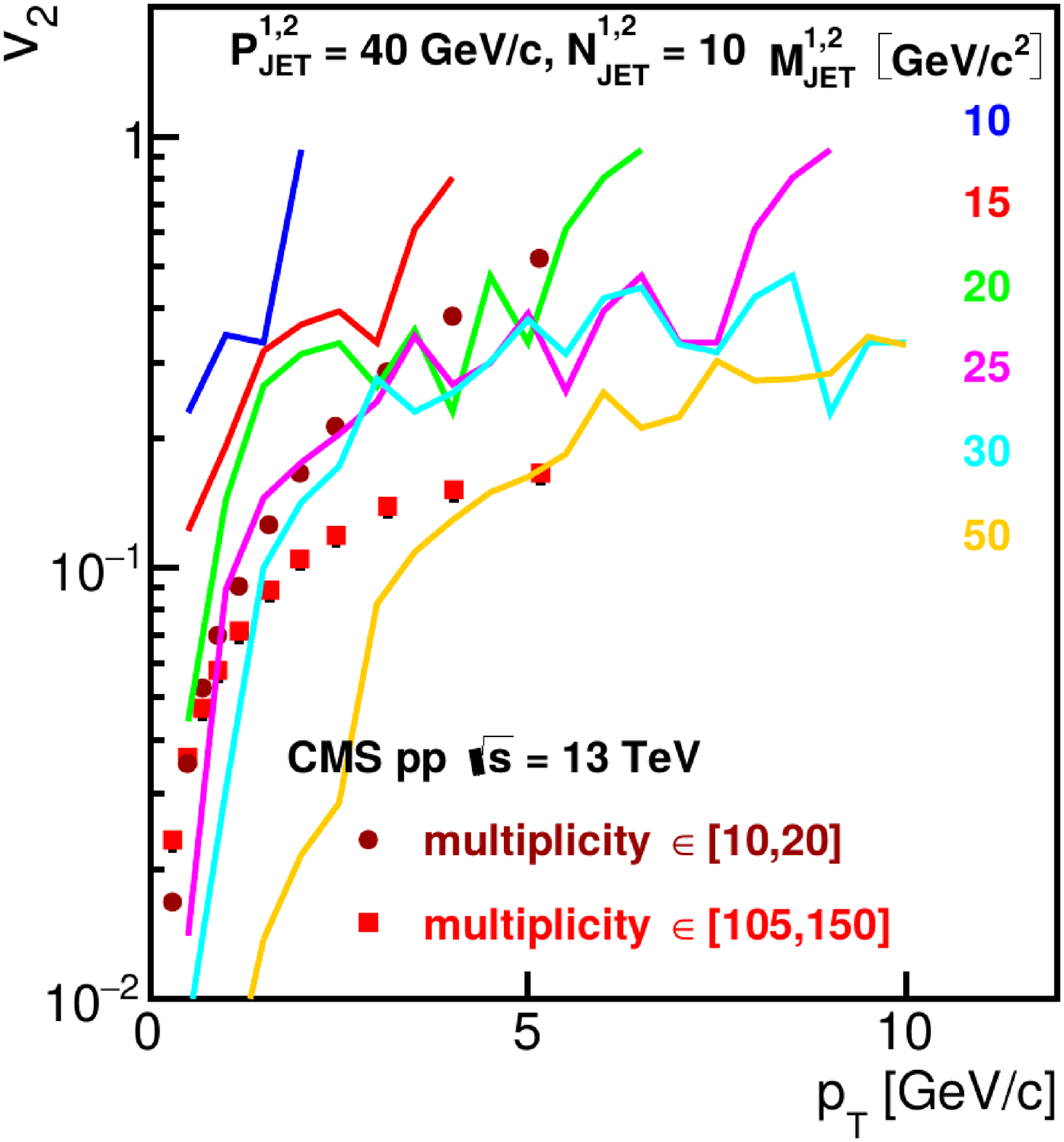} 
\includegraphics[width=0.48\textwidth]{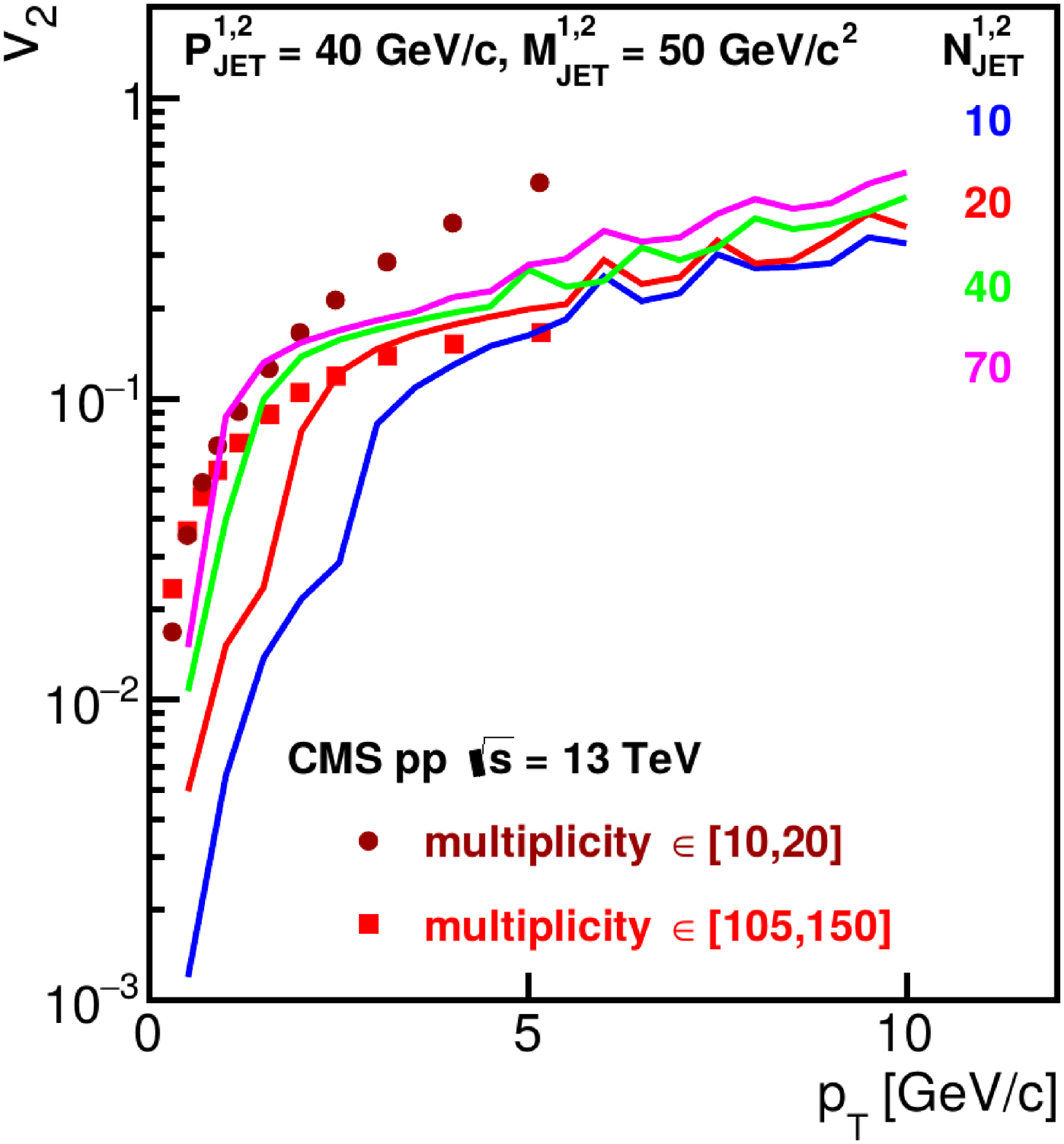} 
\includegraphics[width=0.48\textwidth]{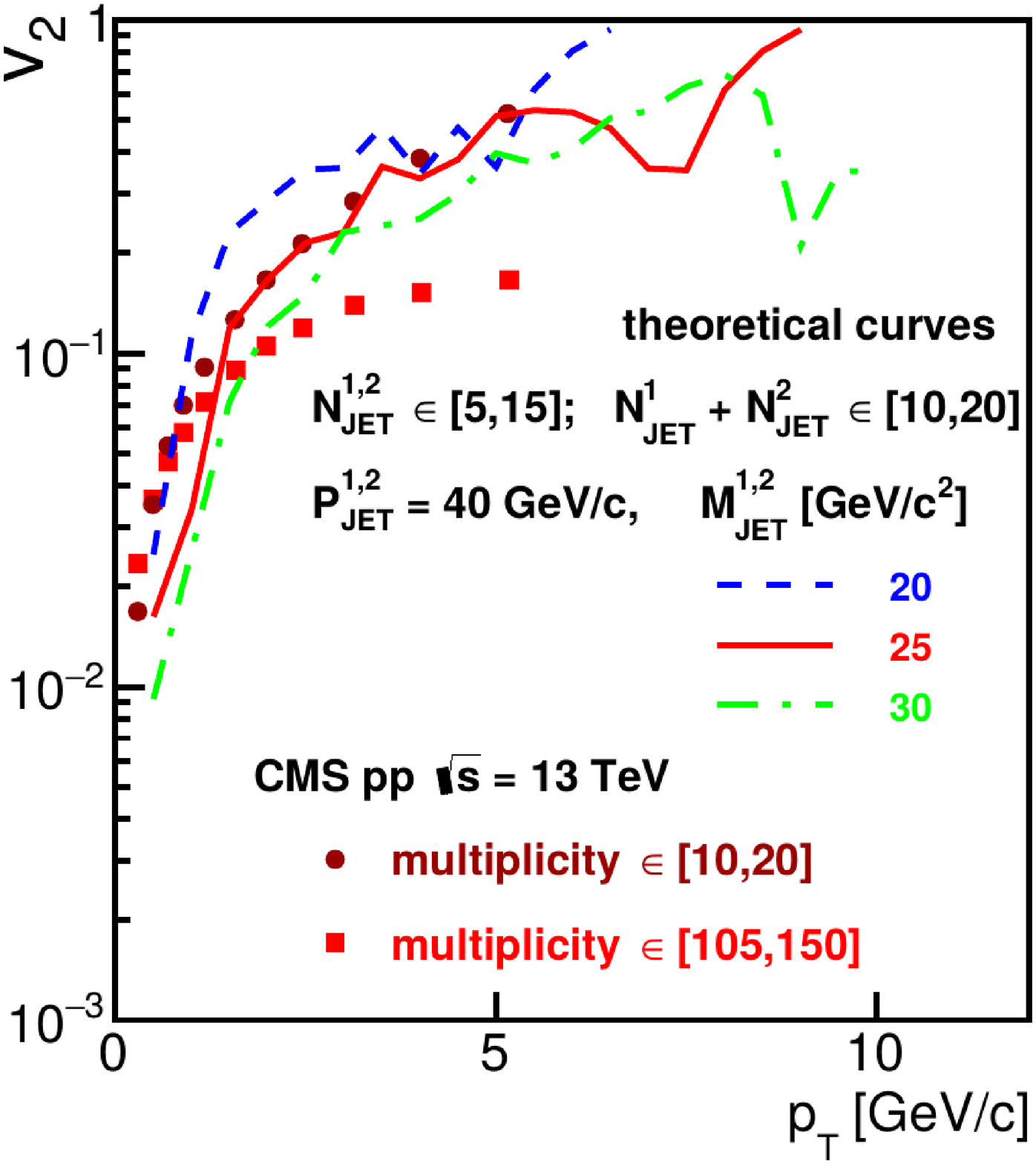} 
\end{center}
\caption{$v_2$ as a function of $p_T$ obtained from long-range ($|\Delta y| \geq 2$) $\Delta y-\Delta\varphi$ correlations. Theoretical curves are obtained using an off-shell fragmentation model for a two-jet hadronic final state with back-to-back jets of various masses and hadron multiplicities. Experimental data are from \cite{bib:CMSv2pp}.
\label{fig:v2}}
\end{figure}

The \textbf{top-left pannel} of Fig.~\ref{fig:v2} shows $v_2$ in cases when the two jets have identical masses, and contain identical numbers of hadrons (10 hadrons per jet, 20 hadrons in total). As the value of jet masses is varied, it can be seen that the bigger the jet mass, the slower the rise of the $v_2$ curve, and the larger the $p_T$ interval, in which, $v_2$ is non-zero. That is, given a sufficiently large jet mass, we have particles with large-enough rapidity separation, yielding non-trivial (non-zero) correlations as well as $v_2$.

In case of the \textbf{top-right pannel} of Fig.~\ref{fig:v2}, the masses and hadron multiplicities of jets are again identical. The jet masses are kept fix at a value ($M_{JET}=50$ GeV/$c^2$) high-enough to produce a non-zero $v_2$ for a large $p_T$ interval, and in the meanwhile, the number of hadrons $N_{JET}$ in the jets is varied. It can be seen that an increment of $N_{JET}$ results in a more steeply rising $v_2$ curve, which is just the opposite of the effect, which we  saw in case of the increase of $M_{JET}$ in the top-left pannel of Fig.~\ref{fig:v2}.

In the \textbf{bottom pannel} of Fig.~\ref{fig:v2}, the masses of the jets are identical, while the hadron numbers in jets are let fluctuate according to the mulciplicity distribution Eq.~(\ref{off5}) (with $r = 33$ and $\bar{p}=0.23$). We use Eqs.~(\ref{cor1})-(\ref{cor2}) to calculate the pair distributions, and let $N$, the total hadron multiplicity in the event vary in the interval of $N \in [10,20]$, just like in the case of the low-multiplicity data set in the experimental analysis \cite{bib:CMSv2pp}. This pannel shows that in case of $M_{JET} = 25$ GeV/$c^2$, the calculated $v_2$ curve is in accordance with the measured low-multiplicity data.

As $v_2$ (from long-range correlations) measured in low-multiplicity pp events can be reproduced via back-to-back jets, the above results support the conjecture that no QGP is created in such cases.

\end{document}